\documentstyle[twoside,epsfig]{article}

\catcode`\@=11
\long\def\@makefntext#1{
\protect\noindent \hbox to 3.2pt {\hskip-.9pt  
$^{{\eightrm\@thefnmark}}$\hfil}#1\hfill}		%CAN BE USED 

\def\@makefnmark{\hbox to 0pt{$^{\@thefnmark}$\hss}}	%ORIGINAL 
	
\def\ps@myheadings{\let\@mkboth\@gobbletwo
\def\@oddhead{\hbox{}
\rightmark\hfil\eightrm\thepage}   
\def\@oddfoot{}\def\@evenhead{\eightrm\thepage\hfil
\leftmark\hbox{}}\def\@evenfoot{}
\def\sectionmark##1{}\def\subsectionmark##1{}}

\oddsidemargin=\evensidemargin
\newcounter{sectionc}\newcounter{subsectionc}\newcounter{subsubsectionc}
\renewcommand{\section}[1] {\vspace{12pt}\addtocounter{sectionc}{1} 
\setcounter{subsectionc}{0}\setcounter{subsubsectionc}{0}\noindent 
	{\tenbf\thesectionc. #1}\par\vspace{5pt}}
\renewcommand{\subsection}[1] {\vspace{12pt}\addtocounter{subsectionc}{1} 
	\setcounter{subsubsectionc}{0}\noindent 
	{\bf\thesectionc.\thesubsectionc. {\kern1pt \bfit #1}}\par\vspace{5pt}}
\renewcommand{\subsubsection}[1] {\vspace{12pt}\addtocounter{subsubsectionc}{1}
	\noindent{\tenrm\thesectionc.\thesubsectionc.\thesubsubsectionc.
	{\kern1pt \tenit #1}}\par\vspace{5pt}}
\newcommand{\nonumsection}[1] {\vspace{12pt}\noindent{\tenbf #1}
	\par\vspace{5pt}}

\newcounter{appendixc}
\newcounter{subappendixc}[appendixc]
\newcounter{subsubappendixc}[subappendixc]
\renewcommand{\thesubappendixc}{\Alph{appendixc}.\arabic{subappendixc}}
\renewcommand{\thesubsubappendixc}
	{\Alph{appendixc}.\arabic{subappendixc}.\arabic{subsubappendixc}}

\renewcommand{\appendix}[1] {\vspace{12pt}
        \refstepcounter{appendixc}
        \setcounter{figure}{0}
        \setcounter{table}{0}
        \setcounter{lemma}{0}
        \setcounter{theorem}{0}
        \setcounter{corollary}{0}
        \setcounter{definition}{0}
        \setcounter{equation}{0}
        \renewcommand{\thefigure}{\Alph{appendixc}.\arabic{figure}}
        \renewcommand{\thetable}{\Alph{appendixc}.\arabic{table}}
        \renewcommand{\theappendixc}{\Alph{appendixc}}
        \renewcommand{\thelemma}{\Alph{appendixc}.\arabic{lemma}}
        \renewcommand{\thetheorem}{\Alph{appendixc}.\arabic{theorem}}
        \renewcommand{\thedefinition}{\Alph{appendixc}.\arabic{definition}}
        \renewcommand{\thecorollary}{\Alph{appendixc}.\arabic{corollary}}
        \renewcommand{\theequation}{\Alph{appendixc}.\arabic{equation}}
        \noindent{\tenbf Appendix#1}\par\vspace{5pt}}
\newcommand{\subappendix}[1] {\vspace{12pt}
        \refstepcounter{subappendixc}
        \noindent{\bf Appendix \thesubappendixc. {\kern1pt \bfit #1}}
	\par\vspace{5pt}}
\newcommand{\subsubappendix}[1] {\vspace{12pt}
        \refstepcounter{subsubappendixc}
        \noindent{\rm Appendix \thesubsubappendixc. {\kern1pt \tenit #1}}
	\par\vspace{5pt}}

\newcommand{\textlineskip}{\baselineskip=13pt}
\newcommand{\smalllineskip}{\baselineskip=10pt}

\def\eightcirc{
\begin{picture}(0,0)
\put(4.4,1.8){\circle{6.5}}
\end{picture}}
\def\eightcopyright{\eightcirc\kern2.7pt\hbox{\eightrm c}} 

\newcommand{\copyrightheading}[1]
	{\vspace*{-2.5cm}\smalllineskip{\flushleft
	{\footnotesize International Journal of Theoretical and Applied Finance#1}\\
	{\footnotesize $\eightcopyright$\, World Scientific Publishing
	 Company}\\
	 }}

\newcommand{\pub}[1]{{\begin{center}\footnotesize\smalllineskip 
	#1\\		%USED ONLY FOR PREPARING THIS INSTRUCTION FILE
	\end{center}
	}}

\def\abstracts#1#2#3{{
	\centering{\begin{minipage}{4.5in}\baselineskip=10pt\footnotesize
	\parindent=0pt #1\par 
	\parindent=15pt #2\par
	\parindent=15pt #3
	\end{minipage}}\par}} 

\renewenvironment{thebibliography}[1]
	{\frenchspacing
	 \ninerm\baselineskip=11pt
	 \begin{list}{\arabic{enumi}.}
        {\usecounter{enumi}\setlength{\parsep}{0pt}     
	 \setlength{\leftmargin 12.7pt}{\rightmargin 0pt} %FOR 1--9 ITEMS
         \setlength{\itemsep}{0pt} \settowidth
	{\labelwidth}{#1.}\sloppy}}{\end{list}}

\newcounter{itemlistc}
\newcounter{romanlistc}
\newcounter{alphlistc}
\newcounter{arabiclistc}

\newcommand{\fcaption}[1]{
        \refstepcounter{figure}
        \setbox\@tempboxa = \hbox{\footnotesize Fig.~\thefigure. #1}
        \ifdim \wd\@tempboxa > 5in
           {\begin{center}
        \parbox{5in}{\footnotesize\smalllineskip Fig.~\thefigure. #1}
            \end{center}}
        \else
             {\begin{center}
             {\footnotesize Fig.~\thefigure. #1}
              \end{center}}
        \fi}

\newcommand{\tcaption}[1]{
        \refstepcounter{table}
        \setbox\@tempboxa = \hbox{\footnotesize Table~\thetable. #1}
        \ifdim \wd\@tempboxa > 5in
           {\begin{center}
        \parbox{5in}{\footnotesize\smalllineskip Table~\thetable. #1}
            \end{center}}
        \else
             {\begin{center}
             {\footnotesize Table~\thetable. #1}
              \end{center}}
        \fi}

\def\@citex[#1]#2{\if@filesw\immediate\write\@auxout
	{\string\citation{#2}}\fi
\def\@citea{}\@cite{\@for\@citeb:=#2\do
	{\@citea\def\@citea{,}\@ifundefined
	{b@\@citeb}{{\bf ?}\@warning
	{Citation `\@citeb' on page \thepage \space undefined}}
	{\csname b@\@citeb\endcsname}}}{#1}}

\newif\if@cghi
\def\cite{\@cghitrue\@ifnextchar [{\@tempswatrue
	\@citex}{\@tempswafalse\@citex[]}}
\def\citelow{\@cghifalse\@ifnextchar [{\@tempswatrue
	\@citex}{\@tempswafalse\@citex[]}}
\def\@cite#1#2{{$\null^{#1}$\if@tempswa\typeout
	{IJCGA warning: optional citation argument 
	ignored: `#2'} \fi}}

\def\pmb#1{\setbox0=\hbox{#1}
	\kern-.025em\copy0\kern-\wd0
	\kern.05em\copy0\kern-\wd0
	\kern-.025em\raise.0433em\box0}

\def\fnt#1#2{\footnotetext{\kern-.3em
	{$^{\mbox{\scriptsize #1}}$}{#2}}}

\def\fpage#1{\begingroup
\voffset=.3in
\thispagestyle{empty}\begin{table}[b]\centerline{\footnotesize #1}
	\end{table}\endgroup}

\def\runninghead#1#2{\pagestyle{myheadings}
\markboth{{\protect\footnotesize\it{\quad #1}}\hfill}
{\hfill{\protect\footnotesize\it{#2\quad}}}}
\font\tenrm=cmr10
\font\tenit=cmti10 
\font\tenbf=cmbx10
\font\bfit=cmbxti10 at 10pt
\font\ninerm=cmr9

\font\eightrm=cmr8

\def\qed{\hbox{${\vcenter{\vbox{			%HOLLOW SQUARE
   \hrule height 0.4pt\hbox{\vrule width 0.4pt height 6pt
   \kern5pt\vrule width 0.4pt}\hrule height 0.4pt}}}$}}

\def\theequation{\thesectionc.\arabic{equation}}	%FOR SETTING EQ.~(1.1)

\begin{document}

\runninghead{A general methodology to price and hedge derivatives 
in incomplete markets}
{A general methodology to price and hedge derivatives 
in incomplete markets}

\normalsize\textlineskip
\thispagestyle{empty}
\setcounter{page}{1}

\copyrightheading{}			%{Vol. 0, No.0 (1992) 000--000}

\vspace*{0.88truein}

\fpage{1}
\centerline{\bf A GENERAL METHODOLOGY TO PRICE}
\vspace*{0.035truein}
\centerline{\bf AND HEDGE DERIVATIVES IN INCOMPLETE MARKETS}
\vspace*{0.37truein}
\centerline{\footnotesize ERIK AURELL}
\vspace*{0.015truein}
\centerline{\footnotesize\it Matematiska Institutionen,
Stockholms Universitet}
\baselineskip=10pt
\centerline{\footnotesize\it S-106 91 Stockholm, Sweden}
\vspace*{10pt}
\centerline{\footnotesize ROBERTO BAVIERA}
\vspace*{0.015truein}
\centerline{\footnotesize\it Dipartimento di Fisica, Universit\`a dell'Aquila,
and Istituto Nazionale Fisica della Materia}
\baselineskip=10pt
\centerline{\footnotesize\it Via Vetoio, I-67010 Coppito, L'Aquila, Italy}
\vspace*{10pt}
\centerline{\footnotesize OLA HAMMARLID}
\vspace*{0.015truein}
\centerline{\footnotesize\it 
Institutionen f{\"or} Matematisk Statistik och F{\"o}rs{\"a}kringsmatematik,
Stockholms Universitet}
\baselineskip=10pt
\centerline{\footnotesize\it S-106 91 Stockholm, Sweden}
\vspace*{10pt}
\centerline{\footnotesize MAURIZIO SERVA}
\vspace*{0.015truein}
\centerline{\footnotesize\it Dipartimento di Matematica, Universit\`a dell'Aquila,
and Istituto Nazionale Fisica della Materia}
\baselineskip=10pt
\centerline{\footnotesize\it Via Vetoio, I-67010 Coppito, L'Aquila, Italy}
\vspace*{10pt}
\centerline{\footnotesize ANGELO VULPIANI}
\vspace*{0.015truein}
\centerline{\footnotesize\it Dipartimento di Fisica, Universit\`a di Roma ``La Sapienza''
and Istituto Nazionale Fisica della Materia}
\baselineskip=10pt
\centerline{\footnotesize\it  P.le A. Moro 2, I-00185 Roma, Italy}
\vspace*{0.225truein}
\pub{ \today }

\vspace*{0.21truein}
\abstracts{
We introduce and discuss a general criterion for
the derivative pricing in the general situation of incomplete markets,
we refer to it as the No Almost Sure Arbitrage Principle.
This approach is based on the theory of optimal strategy 
in repeated multiplicative games originally introduced by Kelly.
As particular cases we obtain the Cox-Ross-Rubinstein and Black-Scholes
in the complete markets case and the Schweizer and Bouchaud-Sornette 
as a quadratic approximation of our prescription.  
Technical and numerical aspects for the practical option pricing,
as large deviation theory approximation and 
Monte Carlo computation are discussed in detail.
}{}{}

\vspace*{1pt}\textlineskip	
\section{Introduction}	
\label{s:introduction}
\vspace*{-0.5pt}
\noindent
The {\it classical} ingredients for the derivative pricing and hedging are 
absence of costs,
efficient and complete market.
According to these hypothesis,
the absence of arbitrage opportunities determines the price.

The most general formulation of the no-arbitrage argument
is due
to Harrison and Kreps~\cite{HarrisonKreps}.
They show that a price system (under certain
restrictions such as no trading costs) admits no arbitrage opportunities
if and only if all price processes are martingales.
This theory is discussed in several
excellent monographs 
~\cite{CoxRubinstein,Duffie,Hull,Ingersoll,Merton}.
Nevertheless the
martingale is not unique
except for a complete market.
In this case it is possible, using options and shares of a 
single stock, to build up a portfolio with deterministic return. Then,
the capital growth rate equals
the bank interest rate.
The most successful (and famous) applications are 
the Black and Scholes formula~\cite{BlackScholes} for continuous time and
the Cox, Ross and Rubinstein for binomial processes~\cite{CoxRossRubinstein}.

Unfortunately, real stock markets are not complete
and a universally accepted pricing procedure is still lacking. 
Nevertheless, the mean-variance methods seem to have the largest
consensus~\cite{BouchaudSornette,Schweizer}.
In these theories, the optimal hedge portfolio
is found by assuming that investors are risk averse. 
One of the main results is that the expected
capital invested in
this portfolio grows according to the bank interest rate.
A serious conceptual and practical problem obviously rises: 
why should a risk averse
investor, or any rational investor, put the money in such a portfolio?
In fact, return is not deterministic and it is expected to 
be merely equal to the fixed bank return.

In the present paper we shall 
consider an incomplete market under the standard assumption
of no transaction costs.
We state an argument which is able to specify 
the appropriate martingale out of the
many possible ones.
Considering the derivative as a possible asset in the diversification of 
a portfolio
we take the point of view of a speculator, who 
is a market operator interested in the ``best'' investment
in the long run limit.
We shall show in section
{\bf 3} in which sense this point of view is reasonable.

The criterion is based on the theory of optimal gambling 
due to Kelly~\cite{Kelly} and hereafter we will refer
to it as the Principle of No Almost Sure Arbitrage. 
Some results of this paper have been already presented 
in~\cite{AurellBavieraHammarlidServaVulpiani} (Aurell et al. 98).

Remark that our methodology differs from the
standard approach, where one takes the point of view
of the investor who tries to minimize risks 
with an appropriate hedging procedure.
In the context of a speculator, we consider in this paper,
the word  ``hedging''  
indicates
the portfolio strategy he chooses to reach his aim.
 
The paper is organized as follows:
in section {\bf 2} we discuss the Kelly theory 
in the general framework of incomplete markets.
In section {\bf 3} 
we state the Principle of No Almost Sure Arbitrage and 
we obtain the derivative price.
In section {\bf 4} we show the {\it almost sure} consequences of
an incorrect pricing.
Section {\bf 5} is devoted to a discussion of
relations and differences with other pricing procedures.
In particular, we obtain the Cox-Ross-Rubinstein and Black-Scholes prices
in the case of complete markets and 
we derive the Schweizer and Bouchaud-Sornette price
as a quadratic approximation of our prescription.
We discuss the large deviation approach in section {\bf 6} and we
develop a Monte Carlo method in section {\bf 7}.
Finally, in section {\bf 8}  we summarize and discuss 
our results.
In the Appendices we discuss in detail some technical points:
the case of time correlated returns
and heavy-tail distributions.

\section{The Kelly theory of Optimal Portfolio}
\label{s:kelly}
\noindent
In this section we summarize  Kelly's theory of optimal gambling~\cite{Kelly}.
Kelly originally was looking for an interpretation of 
Shannon's Information Theory~\cite{Shannon}
outside the context of communication.
Later Breiman~\cite{Breiman1960,BreimanBerkeley}
reconsidered Kelly's theory 
as a model for optimal portfolio in a stock market.
For a recent review of growth-optimal investment strategies
see~\cite{HakansonZiemba,MaslovZhang}.

In the paper we shall consider, except where differently stated,
a discrete time price movement of stock 
(or some other security) described by
\begin{equation} 
S_{t+1}=u_t S_{t}\,\, .
\label{eq:u_i-definition}
\end{equation} 
where time is discrete, $S_t$ is the price at time $t$ 
of a share and the $u_t$'s are independent, identically distributed
(i.i.d. hereafter)
random variables. 

Let us consider a market operator
endowed with a wealth $W_0$ at time zero, 
who decides to gamble on such a stock many times.
He invests at each time
a fraction $l$ of his capital in stock,
and the rest in a risk-less security, i.e. a bank account.
A non-zero risk-less rate
corresponds to a discount factor
in the definition of the share prices, and can be accounted for by
a redefinition of the $u_t$'s.
We set for simplicity the risk-less rate to zero.
We will write the result for the general case 
at the end.
At time $t$ 
the market operator will
hold a number $l W_t/S_t$ of shares, and his wealth
at successive instants of time follows
a multiplicative random process
\begin{equation} 
W_{t+1}=(1-l)W_t + lu_t W_t = \left(1+l(u_t-1)\right)W_t\,\, .
\label{eq:W_i-definition}
\end{equation}
As a consequence
of the large numbers law the exponential growth rate of the wealth is
in the large time limit   a constant
with probability one.
That is,
\begin{equation}
\lambda(l) = \lim_{\tau\to\infty}\frac{1}{\tau} \log\frac{W_\tau}{W_0} 
= E^p[\log\left(1+l(u-1)\right)]
\label{eq:lambda-definition}
\end{equation}
holds for almost all realizations of the
random variables $u_t$, where
\begin{equation}
E^p[\,\, ( \cdot ) \,\,] \equiv \int \,\, ( \cdot ) \,\, p(u) du\,\, .
\end{equation}
and $p(u)$ is the probability density for the random factors $u$.

The optimal gambling strategy of Kelly consists in maximizing
$\lambda(l)$ in (\ref{eq:lambda-definition}) by varying $l$.
We call speculator a market operator who follows this rule.
The solution is unique because the logarithm is a convex function of
its argument.
Let us discuss which
values of $l$ are reasonable in our problem.
First, the optimum $l$ must be such that $1+l(u-1)$ 
is positive on the support of $u$.
Second, we must decide if borrowing of cash
or short selling of stock is allowed.
In the original formulation of Kelly it is not, but here it is useful
to allow $l$ to take any finite
positive or negative value, and look for the maximum of 
$\lambda(l)$.

The desired strategy is specified by the only finite $l^*$ which solves
\begin{equation}
0 = \frac{d \lambda(l)}{dl}|_{l={l}^*} 
= E^p[\frac{u-1}{1+{l}^*(u-1)}]
\label{eq:l-star-definition}
\end{equation}
and the maximum growth rate is
\begin{equation}
\lambda^* = E^p[\log\left(1+{l}^*(u-1)\right)]\,\, .
\label{eq:lambda-star-definition}
\end{equation}

Let us define  
\begin{equation}
q(u) \equiv \frac{p(u)}{1+l^*(u-1)}\,\, ,
\label{eq:q-definition}
\end{equation}
where $l^*$ is the optimal fraction.
It is easy to show that $q(u)$ is a probability density. 
Since
$1+{l}^*(u-1)$ is positive also $q(u)$ is positive.
From (\ref{eq:l-star-definition}) 
and the normalization of $p(u)$ one has 
\begin{equation}
\int  \,\, q(u) du\, = \,\int  \,\, q(u)(1+l^*(u-1)) du = 1 \, \, .
\end{equation}
The $q(u)$ is a new probability with respect to the given stock.
Furthermore, according to this probability the associated 
multiplicative process is a martingale.
From the definition 
\begin{equation}
E^q[\,\, ( \cdot ) \,\,] \equiv \int \,\, ( \cdot )  \,\, q(u) du
\label{eq:Equivalent-Martingale-pricing}
\end{equation}
and using (\ref{eq:l-star-definition}) it is immediate to obtain
\begin{equation}
E^q[u]=1\,\, .
\end{equation}
 
Maximizing any monotonic convex function gives the same formal
result as~(\ref{eq:lambda-definition}) i.e. a unique solution
and, in absence of constraints, the price is a martingale.
However, Kelly has shown that repetition of the investment
many times gives an objective meaning to the statement
that the growth-optimal strategy is the best,
regardless to the subjective attitude to risk or other psychological 
considerations.

The generalization to a speculator who can diversify his investment
on $N$ risk assets is straightforward and the interested reader can find it
in~\cite{Breiman1960}.
In the next section 
we shall consider the particular case of a two risky assets portfolio
composed by a share and a derivative written on the share.
We shall show that the probability~(\ref{eq:q-definition})
plays a privileged role for pricing derivatives.

\section{Principle of No Almost Sure Arbitrage}
\label{s:principle-NASA}
\noindent
We consider now the problem of pricing a
derivative in incomplete markets.
As in the previous section the speculator can invest his capital only at 
discrete times.
 
The speculator's portfolio is composed at time $t=0$ 
by a risk-less security,
a stock and  a derivative written on the same stock
with maturity time $T$ and strike price $K=k S_0$.

At time $T$ the derivative expires and the speculator decides to
buy a new derivative with the same $k$ and maturity $2 T$.  
Let us focus on the investment at time $t$
in order to fix the notation.
We call $C_t$ the value of derivative at time $t$
and  $C_{t+1}(u_t)$ the value at the following instant of time $t+1$.
Finally
\begin{equation}
f_t(u_t) \equiv \frac{C_{t+1}(u_t)}{C_t} \,\, 
\label{eq:derivative-return}
\end{equation}
is the return of the derivative.
The notation $f_{t}(u_t)$ is to stress that we are dealing with
a derivative, i.e. the return of this asset depends
on the return of the share $u_t$. 
The simplest example is an European call option with a strike price $k S_0$,
one time step before 
the expiration date  i.e. $t=T-1$;
in this case one has $C_{T}(u_T-1)=\{u_T-1 S_{T-1}-k S_0\}^{+}$.
 
At time $t$
the speculator diversifies his portfolio by investing 
a fraction $l_t$ of his wealth in shares and a fraction $d_t$ in derivatives.
If the speculator keeps some money $W_S$ in stock and $W_D$ in derivatives,
after the market fall out the two capitals are worth respectively
$W_S u_t$ and  $W_D f_t(u_t)$ where the same random variable $u_t$ 
appears in both expressions.
The total wealth is therefore changed to
\begin{equation}
W_{t+1} = \left(1 + l_t(u_t-1) +
d_t(f_t(u_t)-1)\right)W_{t}\,\, .
\label{eq:double-portfolio}
\end{equation}
 
We shall consider hereafter, as in section {\bf 2},
the return $u_t$ an i.i.d. random variable.
This assumption is relaxed in 
appendix {\bf A} where the markovian case is treated.
 
The return of the derivative $f_t$ is a function of $T - t$ if
$0 < t < T$
and of the random variable $u_t$.
Let us stress that if $u_t$ is an i.i.d. random variable the 
returns $f_t$ 
are periodic of period $T$.
Hence the fractions $\{l_t\}_t$ and $\{d_t\}_t$ 
take on only $T$ different values.

We consider the case of a speculator who plays 
the game every time and
repeats his investment many times.
In this situation 
his wealth will almost
surely grow at an exponential rate 
which is
\begin{eqnarray}
	\lambda(\{l_t\}_{t},\{d_t\}_{t};\{f_t\}_t) &=&
	 \lim_{\tau \to \infty} \frac{1}{\tau} \sum_{n=0}^{\tau}
	 \log\left(1+l_n(u_n-1)+d_n(f_n(u_n)-1)\right)= \nonumber \\
	 &=& \frac{1}{T} \sum_{t=0}^{T -1}
	E^p \left[\log\left(1+l_t(u_t-1)+d_t(f_t(u_t)-1)\right)\right]\,\, .
\label{eq:double-Lyapunov-exponent}
\end{eqnarray}
The speculator is then interested in the fractions 
attaining the optimal growth rate of the capital
$(\{l^*_t\}_{t=0...T-1},\{d^*_t\}_{t=0...T-1})$.
This set of values is unique because of the convexity of the logarithm
given the set of the derivative returns $\{f_t\}_{t=0...T-1}$.

We are now ready to state the Principle of No Almost Sure Arbitrage
(hereafter NASA)
introduced in~\cite{AurellBavieraHammarlidServaVulpiani}:
{\it prices of derivative securities must be such that a speculator 
cannot construct a portfolio out of combinations of
the derivatives and the underlying security which grows almost surely
at a faster exponential rate than a portfolio containing only the
underlying security}.

In other words, the NASA Principle says that the 
returns $\{f_t\}_t$ must be such that the global maximum of
(\ref{eq:double-Lyapunov-exponent}) equals $\lambda^*$, i.e.
\begin{equation}
\lambda(\{l^*_t\}_{t},\{d^*_t\}_{t};\{f_t\}_t)=\lambda^*
\label{eq:master-equation}
\end{equation}
where $\lambda^*$ is the Kelly optimal rate of 
equation (\ref{eq:lambda-star-definition}).
Since one trivially has
$\lambda(\{l_t=l^*\}_{t},\{d_t=0\}_{t};\{f_t\}_t)=\lambda^*$,
the maximum must be in  $(l_t^*=l^*, d_t^*=0)$, 
because of the uniqueness of the maximum of a convex function.

Let us give a simple intuitive interpretation of this fact.
If $d_t^*$ is larger than zero all speculators
would like to buy the derivative in order to achieve 
a larger exponential rate of their capital. 
As a consequence the derivative price rises and
its return rate falls. The resulting fraction
$d_t^*$ decreases. On the other hand, if $d_t^*$ is
less than zero, then all speculators want to go short of 
derivatives causing their price to fall and the return rate to
rise so that $d_t^*$ tends to grow.

From a technical point of view,
the idea of setting $d_t^*$ to zero, is closely related
to the Samuelson's methodology~\cite{SamuelsonWARRANTS} of warrant pricing 
in the `incipient case', see~\cite{Merton}[chapter~7].
We apply such a principle here to derivatives and
not only to warrants.
Furthermore, the condition $d_t^*=0$ stems from a general principle.
The incipient case is 
intrinsically important and not only
a convenient assumption to
simplify the computations.

The condition for $l_t^*$ is
\begin{equation}
0\, =\, \frac{\partial \lambda(\{l_t\}_t,\{d_t\}_t;\{f_t\}_t) }
{\partial l_t}|_{\{l_t=l^*\}_t,\{d_t=0\}_t}\, =\,
 E^p[\frac{u-1}{1+l^*(u-1)}] \,\, .
\label{eq:l-option-star-definition}
\end{equation}
This equation is identical to 
(\ref{eq:l-star-definition}) of the Kelly model and 
we find again the probability $q$ (\ref{eq:q-definition}).
The equation specifying $d^*_t$ is
\begin{equation}
0\, =\, \frac{\partial \lambda(\{l_t\}_t,\{d_t\}_t;\{f_t\}_t) }
{\partial d_t}|_{\{l_t=l^*\}_t,\{d_t=0\}_t}\, =\,
 E^p[\frac{(f_t(u_t)-1)}{1+l^*(u_t-1)}]\,\, .
\label{eq:C_t-star-definition}
\end{equation}
Then using (\ref{eq:q-definition}) 
\begin{equation}
C_t =  E^q[ C_{t+1}(u)] \,\, ,
\label{eq:equivalent-Martingale-pricing}
\end{equation}
which states that the NASA price for derivatives is a martingale
with respect to $q$: in the following we refer to it as the
equilibrium price.
In appendix {\bf B}
we deal with some simple cases where the $q$ 
can be computed analytically,
and we show that even a $p(u)$ very
wildly divergent for large $u$ will lead to
a $q(u)$, which has at least finite first and second
moments. 

We have derived the rule which gives the price at time $t$ given 
the price a time $t+1$.
In particular, the option price at time $T-1$ is
\begin{equation}
C_{T-1}(S_{T-1}) = E^q[C_T(u \, S_{T-1})] \,\, .
\label{eq:lasttime-Martingale-pricing}
\end{equation}  
since the price of the derivative is
known at the expiration time $T$.
If we iterate backwards the above equation
we obtain 
\begin{equation}
C_t(S_t) =  E^Q \left[ C_T(U_{T-t} \, S_t) \right]
\label{eq:equivalent-Martingale-pricing-T}
\end{equation}
where the expectation value $E^Q [\cdot]$ is taken with respect to
the random variable
\begin{equation}
U_{T-t} \equiv \frac{S_T}{S_t} =\prod_{s=t+1}^T u_s
\label{eq:product-u}
\end{equation}
which is the product of $T-t$ 
independent identically $q$-distributed variables $u_s$.
The probability distribution $Q_{T-t} (S_T|S_t)$ of the product 
can be obtained as the convolution of $T-t$ probability distributions
of a single variable.
The price depends on the actual share price $S_t$ at time $t$.
In the particular case 
\begin{equation}
C_T(S_T) = \{ S_T-k S_0 \}^+
\label{eq:call-option-price-final}
\end{equation}
we obtain
\begin{equation}
C_t(S_t) =  E^Q \left[ \left\{ U_{T-t} S_t-k S_0 
\right\}^+ \right]\,\, ,
\label{eq:equivalent-Martingale-pricing-call}
\end{equation}
which is the NASA price of an European call option.

We have taken the risk-less rate to be zero.
A non-zero risk-less rate can be reintroduced as a discount factor
in the share prices.
Actually, having defined $\tilde{u}$ as $u/r$
we can reproduce all the computations for the new stochastic variable.
We start from the effective probability of the $\tilde{u}$ and
we find the probability $q(\tilde{u})$.
Then, for $r > 1$ the European call option price is 
\begin{equation}
C_t(S_t) = E^Q \left[ \left\{ \tilde{U}_{T-t} S_t
- \frac{k S_0}{r^{T-t}} \right\}^+ \right]
\label{eq:equivalent-Martingale-pricing-interest}
\end{equation}
where $ \tilde{U}_{T-t} = \prod_{s=t+1}^T \tilde{u}_s$.

Before we turn to the consequences of a derivative price different 
from the one imposed by the NASA Principle, let us pause and comment.
We are interested in the derivative value 
assuming that the speculator knows but cannot change 
the distribution of stock price returns.
This is of course an idealization but it looks a reasonable starting
point.
We stress once again that the Principle,
as it is stated above, 
can be applied
{\it only} to an asset and a derivative written on it, 
because both of them use the same source of information:
the distribution of the returns of the underlying.
Surely it cannot be applied to determine the stock price.
In this case two different financial objects are involved, 
a risk-free interest rate and a risky asset.
It is not reasonable to have the same growth rate in both cases,
otherwise nobody would invest in the second one. 
However we expect to observe in liquid markets, 
the tendency of 
the average discounted return to become closer and closer to zero.  
This is true for example in the case of currency markets.
In fact after 1989 
for the currencies of the most developed countries
the purchasing power parity is well verified,
i.e. the exchange rate of two countries 
should adjust according to relative prices
\footnote{The observed deviation from this behavior 
is generally due to macroeconomic reasons, such as the rapid 
depreciation of Italian Lira in the wake of ERM crisis in 1992 or
the large appreciation of UK exchange rate in the eighties with the 
development of the North Sea oil fields~\cite{exchange}.}.
A null average return, for periods of one or few years, is 
a simple consequence of the purchasing power parity and it implies
that the optimal fraction in a stationary policy is zero.  

\section{Arbitrage in non equilibrium case}
\label{s:arbitrage}
\noindent
We show here that an incorrect pricing of a derivative 
allows for {\it almost sure} arbitrage i.e. a rate of capital grow
larger than the one obtained with the optimal portfolio 
of shares and risk-less securities.  

Let us suppose that the derivative is not correctly priced
(i.e. it is not at its equilibrium value $C^*_t$):
\begin{eqnarray}
C_t = C^*_t +\Delta C_t & with & t=0,...,T -1 \,\, .
\end{eqnarray}

Given the price of the derivative the problem is to find 
the optimal growth rate of the capital for a mixed portfolio
composed by a risk-less security,
a stock and the derivative written on it.
Assuming that $\Delta C_t$ is small and  expanding the growth rate 
up to the second order in $\delta l_t \equiv l_t-l^*$, $\delta d_t \equiv d_t$ 
and $\Delta C_t$
we find that the optimal portfolio
corresponds to $\delta l_t^*$ and $\delta d_t^*$ given by 

\begin{equation}
\delta l_t^* = \frac{\Delta C_t}{C_t^*}
\frac{\Gamma^{(t)}_{12}}{{\rm det}\Gamma^{(t)}}
\label{eq:deltalstar}
\end{equation}
\begin{equation}
\delta d_t^* = -\frac{\Delta C_t}{C_t^*}
\frac{\Gamma^{(t)}_{11}}{{\rm det}\Gamma^{(t)}}
\label{eq:deltadstar}
\end{equation}
where the matrix $\Gamma^{(t)}$ is
\begin{equation}
\Gamma^{(t)}= \left( 
\begin{array}{cc}
E^p \left[ \frac{(u-1)^2}{(1+l^*(u-1))^2} \right]                       & 
E^p \left[ \frac{(f_t(u_t)-1)(u_t-1)}{(1+l^*(u_t-1))^2} \right] \\
E^p \left[ \frac{(f_t(u_t)-1)(u_t-1)}{(1+l^*(u_t-1))^2} \right] & 
E^p \left[ \frac{(f_t(u_t)-1)^2}{(1+l^*(u_t-1))^2} \right]
\end{array}
\right)  \,\, .
\label{eq:gamma}
\end{equation}

The corresponding maximal rate is
\begin{equation}
\lambda(\{C_t\}_t) = \lambda^* +  \frac{1}{2 T} 
\sum_{t=0}^{T -1}
\left( \frac{\Delta C_t}{C_t^*} \right)^2  
\frac{ \Gamma^{(t)}_{11} } {{\rm det}\Gamma^{(t)}}\,\, .
\label{eq:deltalambdastar}
\end{equation}
Remark that $\Gamma^{(t)}_{11}$ and ${\rm det}\Gamma^{(t)}$ are  
positive quantities.
Therefore, the correction to the equilibrium 
optimal growth rate $\lambda^*$ is always positive 
both for under-valued and over-valued derivatives.

\section{Comparisons with other approaches and limiting cases}
\label{s:comparisons}
\noindent
In this section we compare our approach with previous 
pricing procedures both in case of complete and incomplete markets.

\subsection{Classical pricing prescriptions
for complete markets}
\noindent
The above proposed pricing procedure agrees with no-arbitrage pricing 
in the case of complete markets.
It is, therefore, instructive to carry through the calculations
for the dichotomic Cox-Ross-Rubinstein case and the Black-Scholes 
continuous time limit. 

Following Cox-Ross-Rubinstein we assume that the share 
price can go up by a factor $u_u$ with probability $p$ and 
down by a factor $u_d$ with probability $(1-p)$.
We can safely assume that the risk-free interest rate $r$ equals 
unity. The case $r > 1$
can always be recovered by simply replacing $u$ with $\tilde{u} = u/r$.  
The probability density is therefore:
\begin{equation}
p(u) = p \delta(u-u_u) +(1-p)\delta (u-u_d)
\label{eq:double-delta} 
\end{equation}
where the $\delta(\cdot)$ are Dirac delta functions.

The optimization equation is then solved as
\begin{equation}
l^*= \frac{p}{1-u_d} - \frac{(1-p)}{u_u-1}
\end{equation}
which leads to a probability density $q(u)$ having 
exactly the same form as (\ref{eq:double-delta}),
the only difference being that the probability $p$ 
is replaced by $q$: 
\begin{equation}
q=\frac{p}{1+l_s^*(u_u-1)}
= \frac{1-u_d}{u_u-u_d}
\end{equation}
which is the Cox-Ross-Rubinstein probability $q$.
The new probability $q$ is independent of $p$:
a result which suggests that this case is
somehow atypical.

Consider now the continuous limit case of Black and Scholes.
The prices of shares are continuously
monitored and the financial operator is allowed for 
a continuous time hedging.
In an infinitesimal time interval $dt$
the share changes by a factor
\begin{equation}
u = \exp\{ \eta dt + \Delta dw \}
\end{equation}
where $\eta$ and $\Delta$ are constants
while $dw$ is a random increment with vanishing average and
variance $E^p[(dw)^2]=dt$.
This infinitesimal time expansion is equivalent to say that
\begin{equation}
E^p[\log u]= \eta dt \quad ,  \quad
Var^p[\log u] = \Delta^2 dt
\end{equation}
while it is assumed that higher moments of $\log(u)$
have expected value of higher order in $dt$.

With the above assumptions one can easily solve the equation for $l^*$,
in fact, up to order $dt$ equation for $l^*$  is
\begin{equation}
E^p[(u-1)(1-l^*(u-1))]= 0
\label{eq:eq_lstar_BS}
\end{equation} 
which gives
\begin{equation}
l^* = \frac{\eta +\frac{\Delta^2}{2}}{\Delta^2} 
\label{eq:sol_lstar_BS}
\end{equation}

Using (\ref{eq:sol_lstar_BS}) it is easy to verify that 
\begin{equation}
E^q[\log u]= - \frac{\Delta^2}{2} dt \quad ;  \quad
Var^q[\log u] = \Delta^2 dt
\label{eq:eta-delta}
\end{equation}
and therefore for a finite time $T$  
the variable $U_T = S_T/S_0 $ is distributed according to
a lognormal martingale which 
follows the same prescription of Black and Scholes.

Notice that the continuous case of Black and Scholes refers, as well the 
Cox-Ross-Rubinstein case, to a complete market. 
In both cases it is possible a perfect hedging,
i.e. a complete replication of the option price process by 
a combination of shares and risk-less investments.
This fact is not surprising since the continuous gaussian case
can always be recovered from the dichotomic case,
when the the time step becomes infinitesimal.

It is useful to stress that, 
even if the price 
with our principle is the same of the {\it classical} approaches 
in complete markets, 
this does not mean that the strategy is the same.
The speculator, in our case, does not worry to hedge continuously in order
to obtain a risk-less portfolio but he follows the strategy of trying 
to do the best with the information he has i.e.
he attempts to get the largest {\it almost sure} rate of exponential growth.
Even in the complete market case, the speculator obtains 
an almost sure exponential growth rate $\lambda$ 
generally larger than the risk-less 
interest rate $r$ (obtained by the classical pricing prescriptions), 
explaining why one should invest in
a portfolio which needs an active and frequent trade.

\subsection{ Schweizer and Bouchaud-Sornette 
approach as quadratic approximation}
\label{s:quadratic-approx}
\noindent

In the Schweizer~\cite{Schweizer} and 
Bouchaud-Sornette~\cite{BouchaudSornette}
approach the option price 
is obtained searching an hedging strategy which minimizes  
a quadratic risk function.
This is a natural generalization of the Black and Scholes idea 
of looking for a zero risk strategy (perfect hedging) to the case
in which this is not anymore possible (incomplete markets).
In the limit of continuous time 
the usual Black-Scholes theory is recovered.  
The problem is that the above approach gives
negative values of the option price for large 
$S_T$~\cite{Hammarlid,Wolczynska} 
as shown in a different context by Dybvig and Ingersoll~\cite{DybvigIngersoll}.
Nevertheless, negative prices appear in very unrealistic situations 
while, for more realistic cases, the formula gives sensible values.

We show in this section that the Schweizer and 
Bouchaud-Sornette price can be considered
a quadratic approximation of the NASA one.

Let us first recall their result in a convenient form.
It has been shown in~\cite{Wolczynska} that their price 
can be expressed in the same way of equation
(\ref{eq:equivalent-Martingale-pricing-T})
where the expectation is now constructed using the 
pseudo-martingale probability 
\begin{equation}
q(u)=p(u) \{1- \frac{\mu}{\sigma^2}[(u-1)-\mu] \} \;\; 
\label{eq:mart Wol}
\end{equation} 
where $\mu \equiv E^p[u-1]$, $\sigma^2 \equiv Var^p[u-1]$.

According to a rather general theory,
one can show~\cite{DemangeRochet} 
that the martingale probability associated to
an utility function $U(W)$ in a two times problem is
\begin{equation}
q(u)=p(u)\frac{ \frac{d}{dW}U(W)|_{l=l^*}}
{E^p \left[\frac{d}{dW}U(W)|_{l=l^*} \right]}\,\, .
\label{eq:martingale-theorem}
\end{equation}
where 
\begin{equation}
W=\left(1+l(u-1)\right)W_0\,\, 
\label{eq:due-tempi}
\end{equation}  
is the capital after the investment and $W_0$ before it;
$l^*$ is the optimal fraction associated to the utility $U$.

From equation (\ref{eq:martingale-theorem}) 
is clear why negative prices appear: 
this happens 
only when the utility
is a decreasing function of the capital.

Let us introduce 
a quadratic utility function of the capital $W$
\begin{equation}
U^{quad}(W)=\alpha W - \frac{\beta}{2}(W-\overline{W})^2
\label{eq:utilita-quadratica}
\end{equation}
where $\overline{W}$ is the expected capital and
$\alpha$ and $\beta$ two positive parameters.

Inserting this utility $U^{quad}$ in equation (\ref{eq:martingale-theorem})  
one gets
\begin{equation}
q(u)=p(u) \{ 1 - l^*W_0 \frac{\beta}{\alpha}[(u-1)-\mu]  \} \;\; .
\label{eq:martingale-quad}
\end{equation}
The optimal $l^*$ can be derived simply by the imposition 
of the martingale property or equivalently from the equation
\begin{equation}
l^* W_0 \frac{\beta}{\alpha}=\frac{\mu}{\sigma^2} \;\; ,
\end{equation}
obtaining that the above pseudo-probability coincides
with (\ref{eq:mart Wol}).

Notice that  $ U^{quad}(W) $ is nothing but
a Taylor expansion up to the second order of
a generic utility function around $\overline{W}$
and then also of the logarithmic case 
of the speculator we have considered.
Our price and the Schweizer and 
Bouchaud-Sornette one
coincide for all practical purposes  
in {\it mild} periods
but can lead to different results
if strong fluctuations of the returns are present
(i.e. near financial crisis).

\section{Large deviations}
\label{s:Large deviations}
\noindent
In section {\bf 3} 
we have shown that the NASA price of a derivative
is determined by a probability distribution $Q_T(S_T|S_0)$
which is related to 
the one-time step equivalent martingale probability $q$ defined 
in (\ref{eq:q-definition}).

The probability $Q_T(S_T|S_0)$ is constructed by compounding the 
one time step probability $q$.
If $u$ can take only a finite number of values $N$,
then $Q_T(S_T|S_0)$ can be written as a multinomial formula, 
a simple generalization of the 
Cox-Ross-Rubinstein case with binomial coefficients.

Unfortunately, if the number of time steps is large,
it is not possible to compute explicitly
the multinomial average which gives the price
because it involves $O(N^T)$ operations, 
and some other
approach must be found.
Therefore we need some analytical and numerical techniques 
for an estimation of the price.
In this section we discuss the large deviation approach,
next section is devoted to the Monte Carlo method.

The natural answer to the problem of a correct approximation
of the price comes from large 
deviation theory.
To state the problem we have to define the proper 
large deviation variable

\begin{equation}
z\equiv{1 \over T} \log S_T/ S_0
\end{equation}
whose probability distribution $\tilde{Q}_T(z)$
is trivially obtained from $Q_T(S_T|S_0)$
as
\begin{equation}
\tilde{Q}_T(z) =  T \, \exp \{ zT \} \, Q_T(\exp \{ zT \} ) \;\; .
\label{eq:distribz}
\end{equation}

In a nutshell, the mathematical essence of the 
large deviations theory~\cite{Varadhan} is 
the existence ( under suitable hypothesis: essentially the finiteness 
of all moments of $\exp(Tz)$) of a convex function $G(z)$ 
(usually called Cramers function) such that 
\begin{equation}
\tilde{Q}_T(z) = \Phi_T(z) \exp \{ -G(z) T \}
\label{eq:Approximation LD1}
\end{equation}
where $\Phi_T(z)$ is subexponential for large T.

It is often reported in literature that the distributions
are heavy-tailed, and therefore it could be doubtful the existence
of Cramer function because the hypothesis is not verified.
However distributions obtained by historical sequences 
always have all finite moments, this is a consequence
of the presence of a natural cut-off which can also be reflected 
at the level of models~\cite{MantegnaStanley}.

Because of the Oseledec theorem~\cite{Oseledec} one has that $z$,
in the limit of very large $T$, is almost surely equal to 
its expected value $\lambda$, 
therefore $G(z)$ has to be strictly positive except for $z=\lambda$
where it vanishes.

In order to find out in practice the Cramer function it is sufficient 
to compute the scaling exponents $L(n)$ for the moments of $S_T$ 

\begin{equation}
L(n)\equiv \lim_{T\to \infty}{1 \over T} \log E^Q[(S_T/S_0)^n] .
\label{eq:lyapunov generalized}
\end{equation}
where 
\begin{equation}
E^Q[(S_T/S_0)^n]=\int \tilde{Q}_T(z) \exp\{zTn\} \, dz \, \,.
\label{eq:lyapunov generalized E^Q}
\end{equation}
in fact, the Cramers function and the $L(n)$
are related via a Legendre transformation:

\begin{equation}
G(z)= max_{n} [n z - L(n)]=z n^*(z) - L(n^*(z))
\end{equation}
where $n^*(z)$ is the value where the maximum is realized.

The lognormal approximation can be recovered keeping
only the first two terms of the Taylor expansion of
$L(n)$ around zero. One obtains
\begin{equation}
G(z)= { 1 \over  {2 \Delta^2} } (z-\lambda)^2
\label{eq:parabolic_approximation}
\end{equation}
where
\begin{equation}
 \lambda \equiv E^q[\log u]\,\, , \,\, \Delta^2  \equiv Var^q[\log u]
\label{eq:lognormal parameters}
\end{equation}
which is a good approximation
only for relatively small fluctuation  of $S_T$
around its typical value $\exp \lambda T$,
In spite of its relevance in many 
fields the lognormal approximation is 
from a mathematical point of view rather peculiar.
This is basically due to the fact its  moments  grow too fast
and therefore the Carleman criterion does not hold
\cite{Carleman}.
A simple way is to consider the correction to the parabolic shape 
of the lognormal approximation is to consider the Taylor expansion
of order $R$ of $L(n)$
\begin{equation}
L(n) \simeq \sum_{j=1}^{R} \lambda_j n^j 
\end{equation}
that implies
\begin{equation}
G(z) \simeq \sum_{j=2}^{R} g_j (z-\lambda)^j 
\end{equation}
where
\begin{equation}
\begin{array}{l}
g_2 = { 1 \over  {2 \lambda_2} }\\
g_3 = -{ \lambda_3 \over  {3 \lambda^3_2} } \\
g_4 = {1 \over { \lambda^4_2}} 
 \left( { \lambda^2_3 \over {2 \lambda_2} } -
   { \lambda_4 \over {4} }\right) \\
g_5 = - {1 \over { \lambda^5_2}}
 \left(  { \lambda^3_3 \over {\lambda^2_2} } -
   { \lambda_4 \lambda_3 \over {\lambda_2} } +
  { \lambda_5 \over {5} } \right)  \\
g_6 = {1 \over { \lambda^6_2}}
 \left( {7 \over 3} { \lambda^4_3 \over {\lambda^3_2} } -
   {7 \over 2} { \lambda^2_3 \lambda_4 \over {\lambda^2_2} } +
  {1 \over {\lambda_2} } 
    \left( { \lambda^2_4 \over 2} +  \lambda_5 \lambda_3 \right) -
  { \lambda_6 \over {6} } \right) \\
\vdots
\end{array}
\end{equation}

\begin{figure}
\vspace*{13pt}
\begin{center}
\mbox{\psfig{file=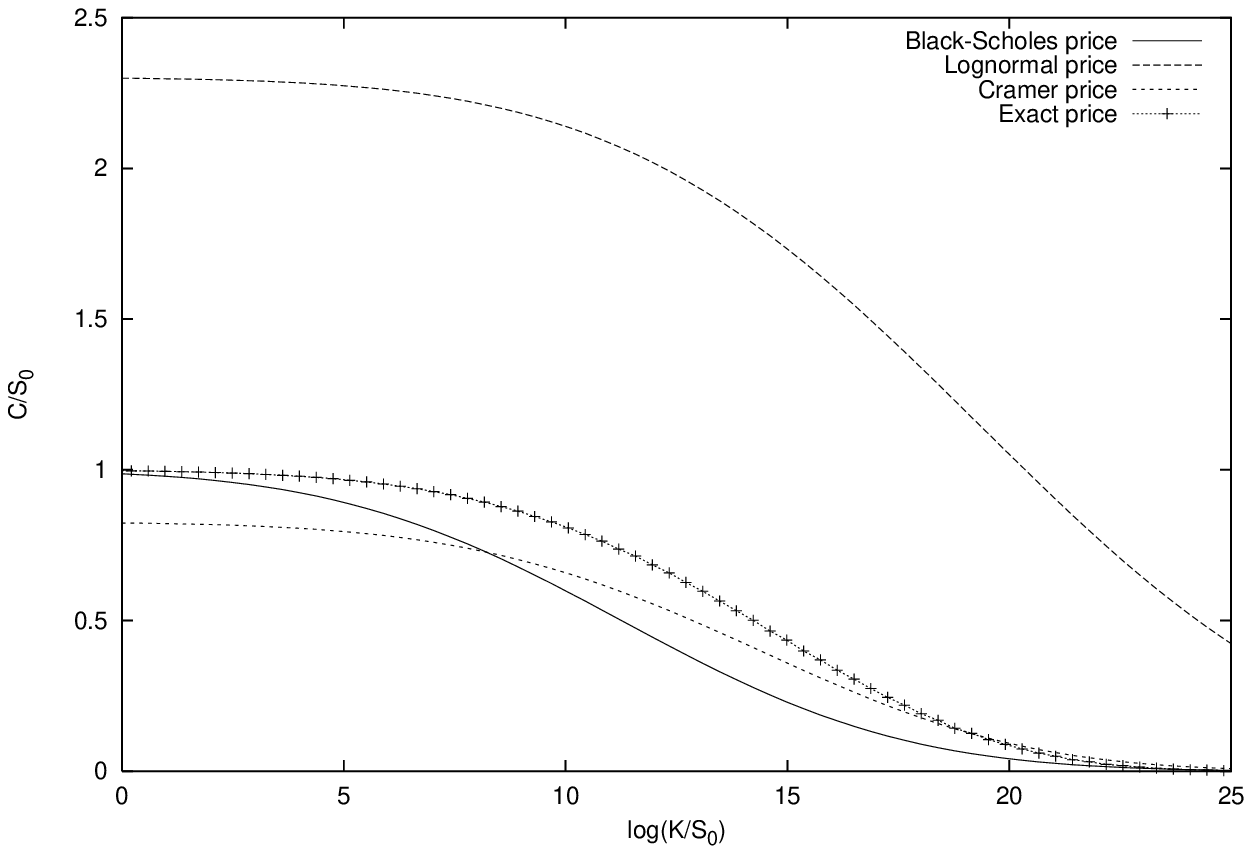,width=4.5in}}
\end{center}
\vspace*{13pt}
\fcaption{
Option prices $C$ rescaled by the initial value of the stock $S_0$
as function of $\log(K/S_0)$  for $T=30$.
The returns of the price over one elementary time step take 
three discrete values $0.2$, $1$ and $2.5$ with probability
$0.15$, $0.15$ and $0.7$ respectively.  
We compare the exact prices, and the Black-Scholes ones  with 
the large deviation prescription with an expansion of $L(n)$ around $0$
with $R=2$ (lognormal approximation) and $R=6$ (see text).
}
\label{fig:Large_zero}

\vspace*{13pt}

\begin{center}
\mbox{\psfig{file=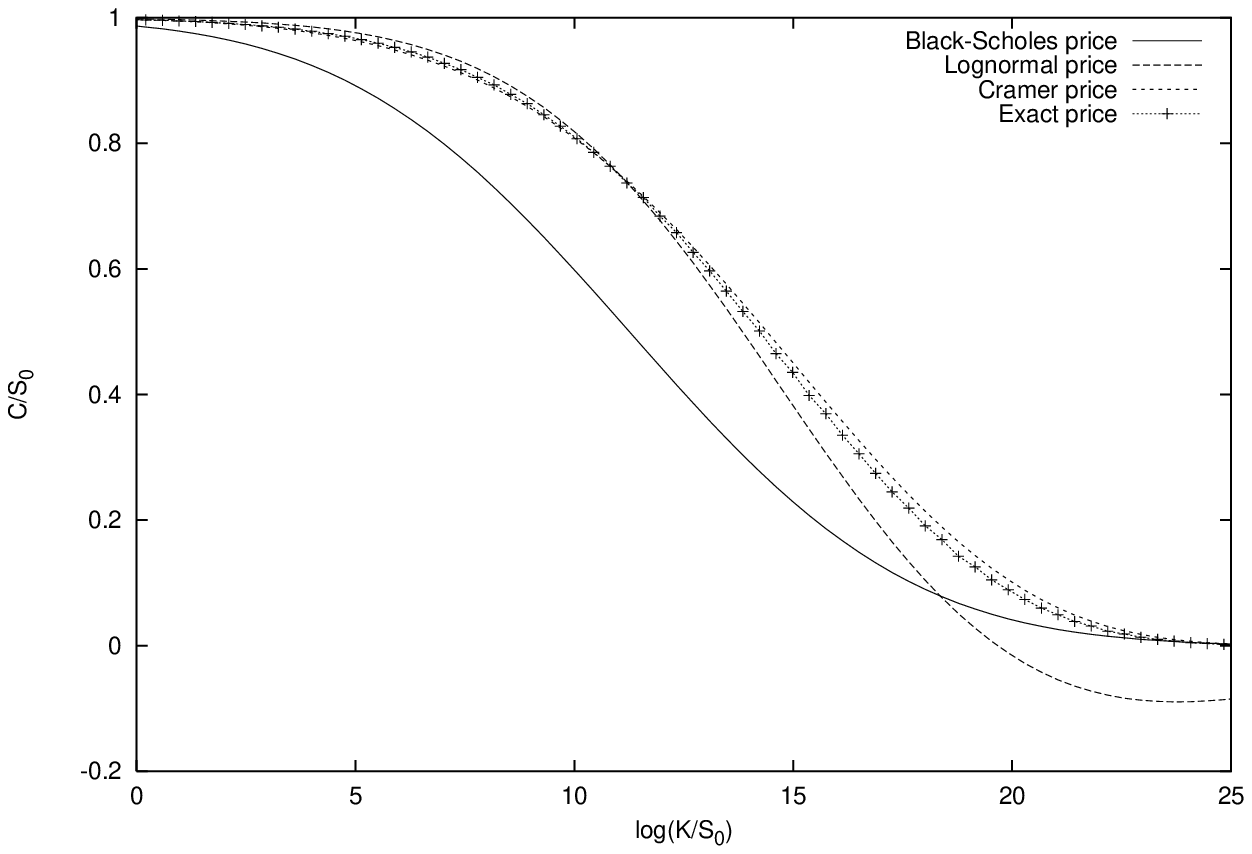,width=4.5in}}
\end{center}
\vspace*{13pt}
\fcaption{
Option prices $C$ rescaled by the initial value of the stock $S_0$
{\it vs.} $\log(K/S_0)$  for $T=30$.
The returns of the price over one elementary time step 
are the same of figure $1$.  
We compare the exact prices, and the Black-Scholes ones  with 
the large deviation prescription described in the text
with $R=2,6$.
}
\label{fig:Large_uno}
\end{figure}

The above expansion of $L(n)$ around $n=0$ 
is quite reasonable because 
in this way one has a rather good approximation of the 
probability distribution
around the maximum.
On the other hand, at least in non trivial cases, one does not obtain
good results for the option pricing, even if
probability normalization factor are properly
taken into account (see figure 
\ref{fig:Large_zero}).

In order to understand this fact it is enlightening to compute 
the scaling exponents $L(n)$ using (\ref{eq:Approximation LD1})
and (\ref{eq:lyapunov generalized}). 
\begin{equation}
E^Q[(S_T/S_0)^n] \sim  
\exp\{ T (nz^*(n) - G(z^*(n))) \}
\label{eq:Scaling exponents zero order}
\end{equation}
where the symbol $\sim$ indicates logarithmic
equivalence and $z^*(n)$ is the maximum of $nz - G(z)$.
The most important contribution to this integral
comes from rare events of exponentially small probability 
such that $z=z^*(n)\neq \lambda$.
Only when $n=0$ 
the integral is dominated by 
the most probable events $z^*(0)=\lambda$.     

Now we can easily see how the above considerations enters into
the problem of pricing options.

Let us notice that the  price $C_0$ of
an European call option can be written as  
\begin{equation}
C_0 = S_0 \int dz \tilde{Q}_T(z) \left\{ \exp T z - k \right\}^{+} =
S_0 I_1  - 
K I_2 
\label{eq:Explicit price LD}
\end{equation}
where
\begin{equation} 
\begin{array}{cc}
 I_1 = \int_{\frac{1}{T} \log k}^{\infty} dz \tilde{Q}_T(z) \exp T z                   & 
 I_2 = \int_{\frac{1}{T} \log k}^{\infty} dz \tilde{Q}_T(z)
\end{array}  \,\, .
\label{eq:integrals}
\end{equation}
Then the origin of the problems both for
the {\it naive} Monte Carlo (see next section) 
and large deviation becomes clear:   
the most important contribution to the integral $I_1$
is given by the exponentially rare events
for which $z=z^*(1)$.
It is then important to observe that because of the martingale property
\begin{equation}
\tilde{\Omega}_T(z) \equiv\tilde{Q}_T(z) exp T z 
\end{equation}
is a probability distribution.
Since the most important contributions to $I_1$
come from the most probable events of the $\tilde{\Omega}_T(z)$ 
distribution,
we can simply repeat the large deviations expansion
of $L(n)$ around $n=0$ for this new probability distribution while
$I_2$ is computed in the previous way. 

It easy to check that $L(n)$
computed with respect to $\tilde{\Omega}_T$
exactly equals $L(n+1)$ computed with respect to $\tilde{Q}_T$.
Therefore the expansion with respect to $n=0$ for
the new probability  distribution corresponds to an expansion 
around $n=1$ for the old one.

All this integrals are not very sensible to 
the choice of the sub-exponential function $\Phi_T(z)$
which can be fixed by means of a normalization constant
or, if more computational precision is needed, as

\begin{equation}
\Phi_T(z) \simeq  \sqrt{T  G^{(2)}(z) \over{2 \pi}} 
\label{eq:Approx fi}
\end{equation}
where $G^{(2)}$ is the $2^{nd}$ derivative of the function $G$.

In figure \ref{fig:Large_uno} we show the results obtained 
in this way for the cases with $R=2$ (lognormal approximation) and 
$R=6$. We observe a quite good agreement with the exact prices. 

As a final remark we want to stress that the lognormal approximation 
does not coincide with the Black and Scholes except in the 
continuous time limit.
The differences can be appreciated in figure 
\ref{fig:Large_zero} and \ref{fig:Large_uno}.

\section{Monte Carlo}
\label{s:Monte Carlo}

\noindent
\begin{figure}
\begin{center}
\mbox{\psfig{file=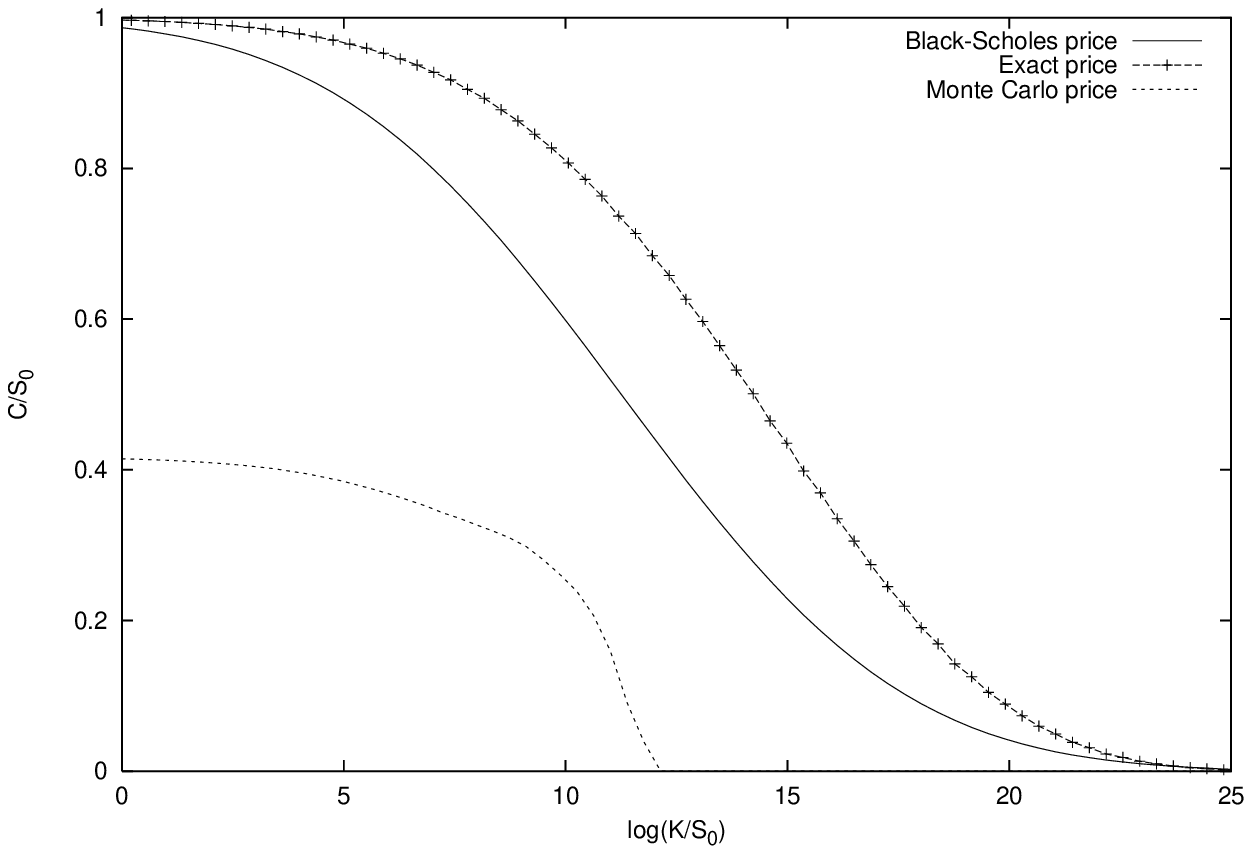,width=4.5in}}
\end{center}
\vspace*{13pt}
\fcaption{
Option prices $C$ rescaled by the initial value of the stock $S_0$
as function of $\log(K/S_0)$ where $K$ is the strike price,
$T$ is chosen equal to $30$.
The returns of the price over one elementary time step 
are the same of figure 1.  
We compare the exact prices, the ones obtained with 
the {\it naive} Monte Carlo algorithm with
$1$ million trials and the Black-Scholes prescription.
The values have been intentionally chosen 
unrealistic in order to stress the differences
between the different approaches.
}
\label{fig:montesemplice}
\end{figure}

The idea to use the Monte Carlo method~\cite{Metropolis}, 
to compute quantities which depend on markovian processes,
is very common 
in both the physicist and economic community.
Instead of computing an average of a function $A(S_T)$ 
according the probability $Q_T(S_T|S_0)$:

\begin{equation}
E^Q[ A(S_T) ]= \int A(S_T) \, Q_T(S_T|S_0) \, d S_T \,\, ,
\end{equation}
one performs a large number $ M $ of trials (i.e. 
$M$ different realizations of the random process $S_T$ 
according to the probability $Q_T$) 
and then compute the average as

\begin{equation}
A_T^{(M)}=
{ 1 \over M} \sum_{j=1}^{M} A(S_T^{(j)})
\end{equation}
where $S_T^{(j)}$ is the $j-th$ realization and 
$E^Q[ A(S_T)]=\lim_{M \to \infty} A_T^{(M)}$.
We discuss the case in which the $u$ can take
only a finite number of values $u^{(i)}$, $i=1,2,...,N$ with probability
$p_i$. The  $j-th$ realization is obtained in the following way.
From  $S_0$ one constructs  $S_1^{(j)}$ as  $S_1^{(j)}=S_0 u_1$,
where  $u_1$ is drawn with 
probability~(\ref{eq:q-definition}),
one extracts in the same way $u_2$ and so on; 
with this procedure one obtains $S_2^{(j)}$,
$S_3^{(j)}$, $\cdots$ and finally  $S_T^{(j)}$. 
In this way the number of operations involved is $O(M T)$, while the 
error estimation using standard variance considerations is

\begin{equation}
\Delta A_T^{(M)} = 
\sqrt{ { 1 \over {M - 1}} \sum_{j=1}^{M} ( A(S_T^{(j)}) - A_T^{(M)} )^2 }
\end{equation}

For example in figure~\ref{fig:montesemplice} 
we show the price of an European call option
as a function of a trinomial probability distribution.
The prices obtained with Monte Carlo simulations with
$1$ million trials are compared with the Black-Scholes ones and 
the exact ones given by Eq.~(\ref{eq:equivalent-Martingale-pricing-call}).

The large errors involved in this approximations are mainly due to the
fact that the most important contribution 
in~(\ref{eq:equivalent-Martingale-pricing-call}) comes from the tails
of the distribution $Q_T(S_T|S_0)$.
We need then a good control of the tails; this is provided 
by the large deviations theory or a suitable Monte Carlo.

\begin{figure}
\vspace*{13pt}
\begin{center}
\mbox{\psfig{file=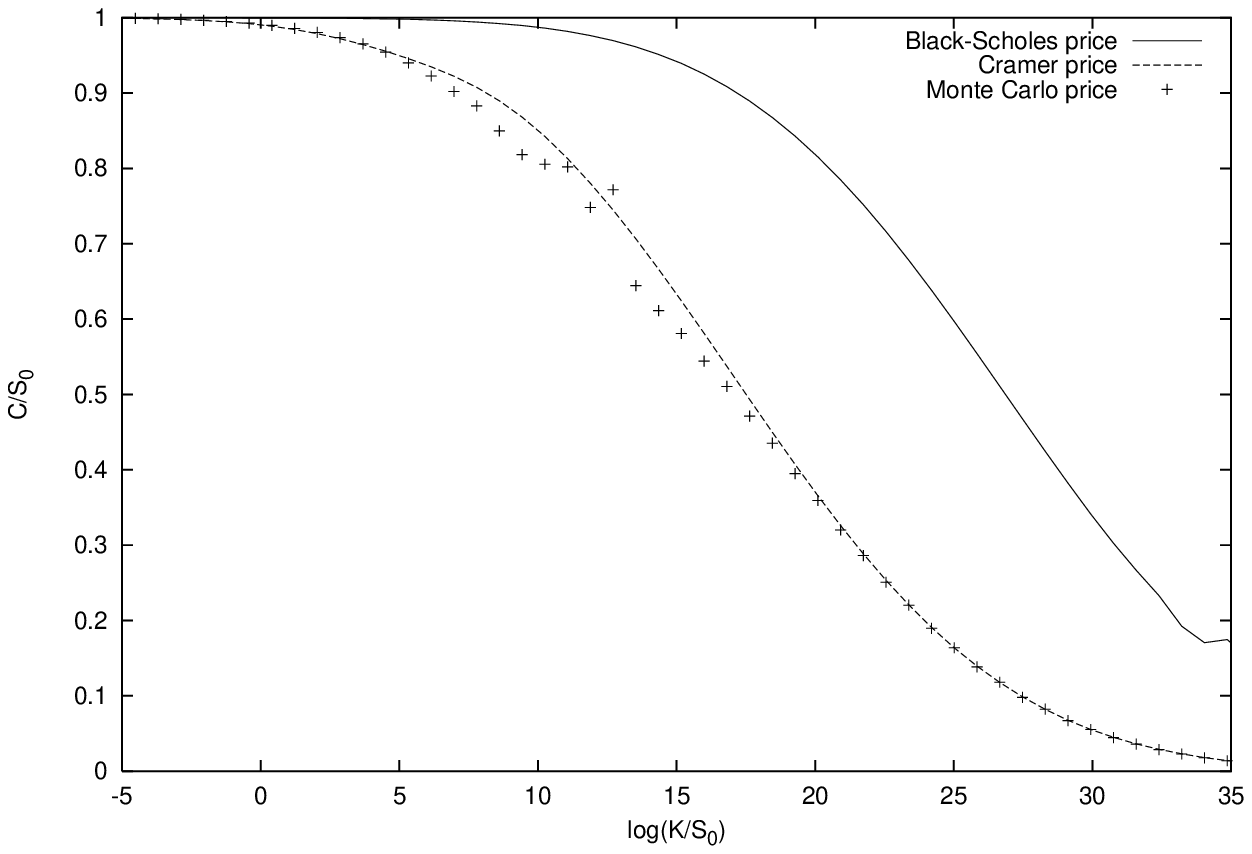,width=4.5in}}
\end{center}
\vspace*{13pt}
\fcaption{
Option prices $C$ rescaled by the initial value of the stock $S_0$
{\it vs.} $\log(K/S_0)$ for $T=30$.
The returns can take the values $u^{(n)} = u^{(0)} \alpha^n$,
with $u^{(0)}$ equal to $0.4$, $\alpha$ equal to $1.2$ and the
index $n$ ranging from zero to $N=30$.
The probability $p(u^{(n)})$ is $C_N (u^{(n)})^{-\beta}$, with
$\beta = 0.5$ and $C_N$ a normalization constant.  
We compare the large deviations ($R=6$) and Monte Carlo
approximations and
the Black-Scholes prices.
The Monte Carlo simulation has been performed with $10^7$ trials.
}
\label{fig:farwest}

\vspace*{13pt}

\begin{center}
\mbox{\psfig{file=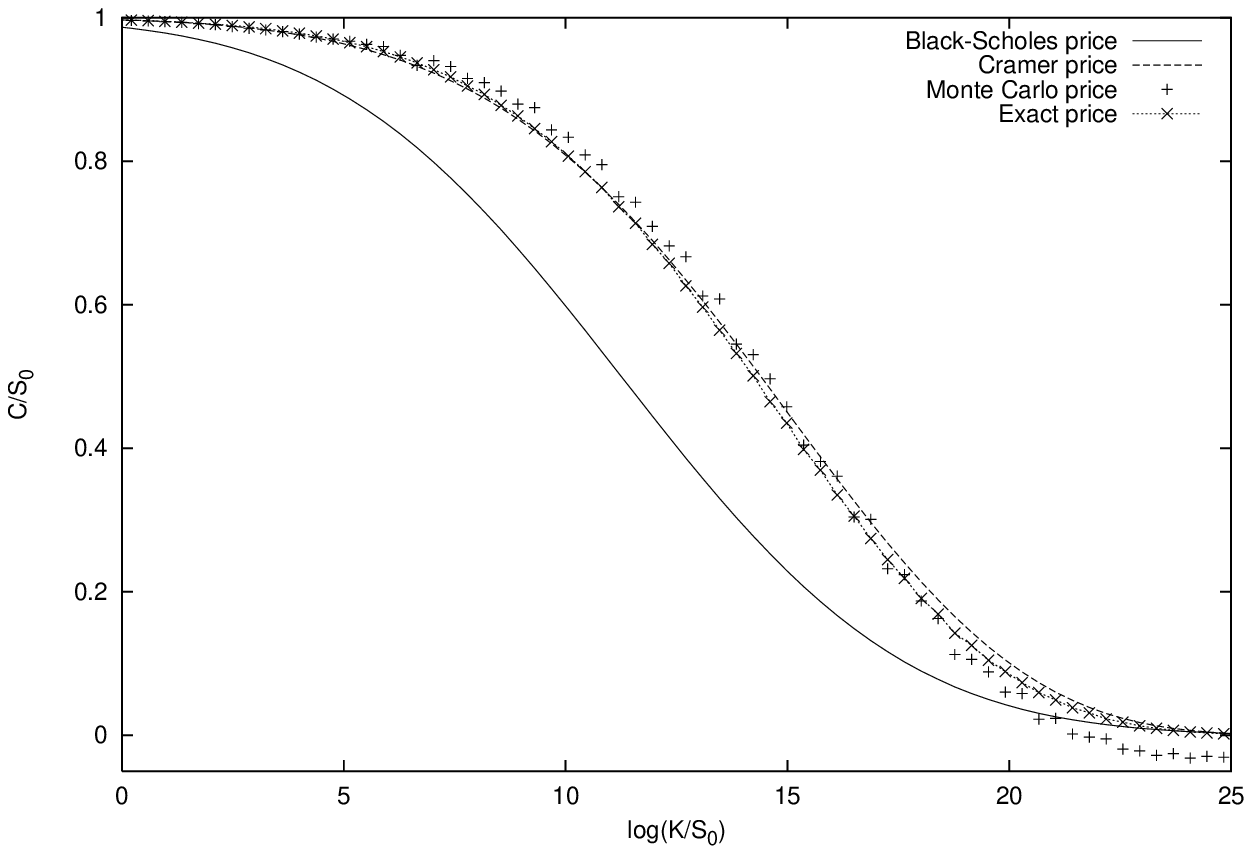,width=4.5in}}
\end{center}
\vspace*{13pt}
\fcaption{
We compare the same quantities of figure~\ref{fig:farwest} 
with the exact prices
in the case of the trinomial returns of figure $1$.  
The Monte Carlo simulation has been performed with $10^5$ trials.
}
\label{fig:Large_all}
\end{figure}

Taking into account the results of the previous section
we have now all the ingredients to build a reasonable Monte Carlo algorithm.
As previously shown in (\ref{eq:Explicit price LD})
the price is a linear combination of two integrals, $I_1$ and $I_2$. 
The most important contribution to $I_1$ come from the most
probable events of $\tilde{\Omega}_T$. Let us write 
\begin{equation} 
 I_1 = \int_{\frac{1}{T} \log k}^{\infty} dz \tilde{\Omega}_T(z) 
= \sum_{S_T > K} \Omega_T(S_T/S_0)  
\label{eq:integral1}
\end{equation}
where $\Omega$, in the case $u$ assumes only a finite
number of values $N$, is the convolution of 
a number $T$ the one-time probability defined by 
\begin{equation} 
\omega(u^{(i)}) = u^{(i)} q(u^{(i)})
\label{eq:omega-definition}
\end{equation}
that is properly normalized because of the martingale 
property.
The algorithm follows then the same steps of the simple Monte Carlo,
using to compute the integral $I_1$ 
the $\omega$-probability instead of the $q$-probability.
To improve the results we have also used the control variate
technique~\cite{Boyle}.
Even in the extreme situation of a very large $Var[u]$ such as 
in the case of a truncated Levy distribution
of figure~\ref{fig:farwest} we obtain good results
both for the Monte Carlo and the large deviations approximation.
In figure~\ref{fig:Large_all} we show the results obtained 
the trinomial distribution of figure 1.

\section{Final Remarks}
\label{s:conclusions}
\noindent
In this paper we have discussed a general criterion to price derivatives.
This principle (No Almost Sure Arbitrage) basically says that it
is not possible to build a portfolio including derivatives which
grows almost surely with an  exponential rate larger than one
with only the 
underlying securities.

This approach holds for the general case of incomplete markets 
and for a generic
probability distribution of the return. 
Let us stress that it is possible to apply the procedure also
for returns correlated in time, e.g. a Markov process.

In the cases of dichotomic distribution
for the return and continuous time limit, 
our method gives the same results of Cox-Ross-Rubinstein and
Black-Scholes  respectively.
If  $p(u)$ is not too spread, the results obtained with 
our approach are rather close to those ones of 
the Schweizer and Bouchaud-Sornette theory.
This is not true in presence of strong fluctuations for the return.
In this case an accurate computation of the option price can be obtained
in the framework of the large deviations theory which systematically takes
into account the deviation from the lognormality or via a properly realized
Monte Carlo computation.

\nonumsection{Acknowledgment}
\noindent

We thank
Patrizia Castiglione and Guglielmo Lacorata  
for helpful discussions about the numerical implementation of the method, 
Lisa L\"{o}fvenberg for a careful reading of the manuscript and 
Paolo Muratore Ginanneschi for useful remarks about large deviations.
We also thank  Jean-Philippe Bouchaud for his helpful comments.
R.B., M.S. and A.V. thank Michele Pasquini for many enlightening 
discussions about Kelly strategies. 
R.B. acknowledges the Royal Institute Technology of Stockholm 
where part of this work has been developed. 

\renewcommand{\theequation}{\Alph{appendixc}.\arabic{equation}}
\appendix{ A}

\noindent
In this appendix 
we consider a Markovian model for the price movement of the stock 
(i.e. equation~(\ref{eq:u_i-definition})) in order to extend
our price prescription to this case. 
Now the $u_t$ are random variables with a Markovian probabilistic 
rule.
For sake of simplicity  we
discuss the case of Markov chain, i.e. $u_t$ can take only $N$
different values $ u^{(1)},  u^{(2)}, \cdots   u^{(N)}$.

Let us remind that a Markov chain is completely characterized by its
transition matrix

\begin{equation}
P_{i\to j}=Prob(u_{t+1}=u^{(j)}|u_{t}=u^{(i)}) \, .
\end{equation}
In a realistic market it is natural  to assume that the Markov chain is 
ergodic and therefore there exist an invariant probability:

\begin{equation}
 p_j=\sum_{i=1}^N p_i P_{i\to j}\,\, .
\end{equation}
For ergodic chain the convergence to the 
invariant probability is exponentially fast

\begin{equation}
(P^n)_{i\to j}=Prob(u_{n}=u^{(j)}|u_{0}=u^{(i)})=
             p_j+O(e^{-\alpha n})
\end{equation}
where $\alpha$ is given by the second eigenvalue of the matrix
$P_{i\to j}$.

The optimization strategy for the shares problem is rather obvious.
If at time $t-1$, $u_{t-1}=u^{(i)}$ then the probability to have  
$u_{t}=u^{(j)}$ is nothing but $P_{i\to j}$. Therefore the optimal
fraction of capital in stock $l_i$, is obtained from the maximization
of

\begin{equation}
\sum_{j=1}^N P_{i\to j} \log [1+l_i(u^{(j)}-1)] \,\, .
\label{eq:markov-NASA}
\end{equation}
Of course  $l_i$ can change at any $i$. In Kelly's paper is discussed
the relation  between the Shannon entropy of the Markov chain and the
optimal exponential growth rate of the capital.

Following the same step of Kelly's strategy described 
in section {\bf 2} it is straightforward to construct
from $P_{i\to j}$ an equivalent martingale with transition probability
$Q_{i\to j}$:

\begin{equation}
Q_{i\to j}={ {P_{i\to j}} \over { 1+l^*_i(u^{(j)}-1) } } \,\, ,
\label{eq:markov-martingale}
\end{equation}
where $l^*_i$ is given by the maximization of the 
quantity~(\ref{eq:markov-NASA}).
It is easy to realize that
\begin{equation}
E^Q(u|u^{(j)})=1 \, \, .
\end{equation}

If the Markov chain with transition matrix $P_{i\to j}$ is ergodic
then also the Markov chain with transaction matrix $Q_{i\to j}$
has to be ergodic and one can introduce an invariant probability

\begin{equation}
 q_j=\sum_{i=1}^N q_i Q_{i\to j} \, \, .
\end{equation}
Also for the new Markov chain the convergence to the 
invariant probability is exponentially fast

\begin{equation}
(Q^n)_{i\to j}= q_j+O(e^{-\beta n}) \,\, ,
\end{equation}
nevertheless, in general, $\beta \ne \alpha$.

Now we have all the ingredients for the construction
of $Q_T(S_t|S_0)$ since we can assign the probability to each trajectory.
Then, by repetition of the same steps of section {\bf 3},
one can construct the price of the derivatives.
The only difference is that now the expectation is taken with 
respect to the markovian martingale.

Let us notice that for the dichotomic case $N=2$, the
present construction
gives the same probabilistic rule of 
Cox-Ross-Rubinstein. In this case the
$Q_{i\to j}$ do not depend on $P_{i\to j}$ but only on the values
$u^{(1)}$ and $u^{(2)}$.
The Monte Carlo computation is rather simple.
In each trial, at time $t$, if $u_{t-1}=u^{(i)}$,
one chooses with 
probability $Q_{i\to j}$
the $j$-th  branch of the tree.

Also the lognormal approximation for the Markovian case is 
straightforward but interesting.
Equation~(\ref{eq:parabolic_approximation}) holds, now one has 

\begin{equation}
\lambda=  \lim_{T\to\infty} { 1 \over T} 
E^Q[\sum_{t=1}^T \log u_t]=E^Q[ \log u ] 
\end{equation}
and 
\begin{equation}
\Delta^2=  \lim_{T \to \infty} { 1 \over T}
 E^Q[(\sum_{t=1}^T (\log u_t-\lambda))^2] \,\, .
\end{equation}
After having defined
\begin{equation}
R_n \equiv
\sum_{i,j} q_i (\log u^{(i)}-\lambda)(\log u^{(j)}-\lambda) (Q^n)_{i \to j}\,\, ,
\end{equation}
it is a matter of computation to check,
using the exponential convergence of $R_n$
due to the Markovian nature of the process,
that
\begin{equation}
\Delta^2=  E^Q[(\log u-\lambda)^2] + 
 2   \sum_{n=1}^{\infty} R_n\,\, .
\label{eq:markov-correlations}
\end{equation}

Let us note the fact that,
at variance with $\lambda$ which depends only on the invariant probability,
for $\Delta^2$   are also relevant the time correlations.
Equation~(\ref{eq:markov-correlations}) 
shows that the "effective" volatility takes into account the time
correlation.

\renewcommand{\theequation}{\Alph{appendixc}.\arabic{equation}}
\appendix{ B}

\noindent
It is useful to have some cases in which 
the probability $q$ is easy to handle analytically.
In this appendix we consider three particular distributions of the 
return of the asset, 
in the case of small  
excess rate of return $\mu \equiv E^p[u-1]$.

\smallskip
{\it Compact support distribution}

\noindent
We first 
consider a probability distribution $p(u)$ in a compact support 
$[0,u_{max}]$.
Equation (\ref{eq:l-star-definition}) can be rewritten as
\begin{equation}
\sum_{n=0}^{\infty} 
E^p [(-1)^n (l^*)^n(u-1)^{n+1}] = 0\,\, .
\label{eq:expanded-denominator}
\end{equation}
We can solve for $l^*$ as a power series in the 
excess rate of return $\mu$

\begin{equation}
l^* = \frac{1}{\sigma^2}\mu + \frac{\chi}{\sigma^6}\mu^2+
O \left(\mu^3 \right) \,\, ,
\label{eq:l-star-compact}
\end{equation}
where $\sigma^2$ and $\chi$ are the second and the third cumulant
of the rescaled return $u-1$ respectively.
From equations~(\ref{eq:q-definition}) and~(\ref{eq:l-star-compact})
we get the probability
\begin{equation}
q(u) = p(u) \left(1-\frac{\mu}{\sigma^2}(u-1)
+\mu^2\left(-\frac{\chi}{\sigma^6}(u-1) + \frac{1}{\sigma^4}(u-1)^2\right)
+ O \left(\mu^3 \right) \right) \,\, ,
\end{equation}
hence up to order $\mu$ the result of the NASA principle is 
identical to the
minimal risk prescription of Schweizer and
Bouchaud-Sornette.

Expansion~(\ref{eq:expanded-denominator}) is convergent if the
optimal value $l^*$ is small enough. 
The individual terms in
the expansion however exist under the much weaker conditions that the
respective moments are finite, but the support of $p(u)$
can be unbounded. 

\smallskip
{\it Levy distribution}

\noindent
There are some known experimental evidences that the one-time  
return $u$ is well described by a proper Levy 
distribution~\cite{Mandelbrot,MantegnaStanley}, 
but it is still an open problem if the variance is finite or not.
In this case expansion~(\ref{eq:expanded-denominator}) is
meaningless.
We assume here that the returns 
are described by a probability distribution with
a power law decay at large arguments
\begin{equation}
p(u) \sim \beta u^{-1-\nu} \, \, \hbox{for} \, \, u>>1 \,\, .
\label{eq:power-law-tail}
\end{equation}
The exponent of the power law $\nu$ is in the interval 
$(1,2]$ and $\beta$ is a constant.
Hence the expected value of $u$ exist and is finite, but all
the higher moments are infinite.

Let us notice that the maximizing $l^*$ cannot be less than zero,
because the support of $p$ is unbounded. Hence, $l=0$ is
at the boundary of the domain for which $\lambda(l)$
is well-defined function of $l$.

Let us now assume that $l^*$ is small, but not necessarily of
the same order as $\mu$,
in this case the probability distribution $q(u)$ is well approximated by
\begin{equation}
q(u)\sim \left\{ \begin{array}{ccc}
	p(u)                 &\hbox{if}& u<< \frac{1}{l^*}\\
	\frac{1}{l^* u} p(u) &\hbox{if}& u>> \frac{1}{l^*}
         \end{array}   \right.  \,\, ,
\label{eq:q-levy}
\end{equation}
and then its second moment exists.
Using the approximation~(\ref{eq:q-levy}) for the probability $q$
and the definition of $\mu$, after some algebra one obtains
\begin{equation}
l^* \approx \left(\frac{\mu\nu(\nu-1)}{\beta}\right)^{\frac{1}{\nu-1}} \,\, .
\label{eq:l^*_equation}
\end{equation}
In the limit when $\nu$ tends to $2$, $l^*$ scales linearly with
$\mu$, in agreement with (\ref{eq:l-star-compact}).
When $\nu$ is in the interval between $1$ and $2$, the optimal fraction
however goes to zero faster than $\mu$. 

\smallskip
{\it Log-Levy distribution}

\noindent
We construct here the probability $q$ from a $p$ of the form
\begin{equation}
p(u) \sim \left\{ \begin{array}{ccc} 
		  C u^{-1}(\log u)^{-1-\nu^{(+)}} & \hbox{for} & u>>1 \\
	  	  C u^{-1}|\log u|^{-1-\nu^{(-)}} & \hbox{for} & u<<1
       	  	\end{array}   \right.   
\label{eq:log-Levy-tail}   
\end{equation}
We will refer to these laws as ``log-Levy'' laws, since
they are distributed as Levy laws with exponents
$\nu^{(+)}$ and
$\nu^{(-)}$
in the logarithmic variable.
In this case the expected value of the logarithm of the return 
is finite but the variance is infinite.
As the Levy laws are stable under addition, so the log-Levy
laws are stable under multiplication,
if the
exponents
 of the power laws,
$\nu^{(+)}$ and
$\nu^{(-)}$, are in the interval
$[1,2]$.
The limit case of an exponent equal to $2$ corresponds to
the lognormal distribution.

However, this does not mean that the log-Levy laws are
necessarily as natural as the Levy laws. The divergence
at large $u$ in (\ref{eq:log-Levy-tail}) implies
that the
expectation value with respect to $p$
 of the random variable $u$ does not exist.
Therefore the 
excess rate of return $\mu$ is undefined.

If the probability distribution $p$ is nicely behaved for
small $u$ we have that the growth rate of the 
wealth (\ref{eq:lambda-definition})
exists and is finite in a region around $l=0$.
We can therefore find the maximizing $l^*$,
attained at some positive $l^*$. 
Using the variational
equation, we obtain the probability (\ref{eq:q-definition})
that can be approximated by~(\ref{eq:q-levy}).
Just as in the Levy distribution,
we have that $E^q[u^2]$ is finite,
and it follows the much weaker condition that
$ E^q[(\log u)^2] $ is convergent at large $u$.

\nonumsection{References}

\end{document}